\documentclass{article}

\PassOptionsToPackage{numbers, compress}{natbib}

\usepackage[preprint]{neurips_2022}

\usepackage[utf8]{inputenc} 
\usepackage[T1]{fontenc}    
\usepackage{hyperref}       
\usepackage{url}            
\usepackage{booktabs}       
\usepackage{amsfonts}       
\usepackage{nicefrac}       
\usepackage{microtype}      
\usepackage{xcolor}         
\usepackage{graphicx}

\title{A Learned Simulation Environment to Model Plant Growth in Indoor Farming}

\author{
J. Amacker $^1$, T. Kleiven $^2$, M. Grigore $^2$, P. Albrecht $^2$, and C. Horn $^1$ \\
$^1$ Zurich University of Applied Sciences, ICLS, Schloss 1, 8820 Wädenswil, Switzerland \\
$^2$ Fruitful Farming AG, Höhenstrasse 11, 8954 Geroldswil, Switzerland 
}

\begin{document}

\maketitle

\begin{abstract}
We developed a simulator to quantify the effect of changes in environmental parameters on plant growth in precision farming. 
Our approach combines the processing of plant images with deep convolutional neural networks (CNN), growth curve modeling, and machine learning.  
As a result, our system is able to predict growth rates based on environmental variables, which opens the door for the development of versatile reinforcement learning agents.
\end{abstract}

\textbf{Keywords:} indoor farming, plant growth modeling, deep reinforcement learning.

\section{Introduction}

Precision farming technologies enable sustainable fresh food production and supply chain by allowing to control and steer the growth environment. Controlling and steering the environment makes it possible to continuously optimize the growth path of every plant. However, depending on external factors, no one-size-fits-all growth recipes exist. Instead, there is a need to continuously monitor plant growth and adapt to the environment based on plant behavior.  

In this research, we monitor the environmental conditions using air and water quality (IoT) sensors, and capture behavior in reaction to changes in the environment. We focus on  plants’ circadian rhythm, leaf area, plant growth, and plant color.   

Our main contributions are: 
\begin{itemize}
    \item 
    An end-to-end approach to plant growth modeling from seed to harvest; 
    \item 
    An efficient hybrid curve-fitting and machine learning approach to quantify the effect of environmental parameters on plant growth; 
    \item 
    A demonstration of how model-based reinforcement learning (RL) can be enabled by learning components of a simulator.  
\end{itemize}

The benefits of this approach translate to reduced operating costs, increased quality of the crop, and stable yield. 

\section{Method}
Our aim is to build a simulator that allows us to quantify the effect of changes in environmental parameters on plant growth. This could then be used to understand and improve the growth process. In particular, such an approach provides the basis for the development of autonomous, intelligent agents to be deployed in indoor farms. Using these agents, one can dynamically optimize environment parameters for every plant as a function of its current conditions.  
\begin{figure}
  \begin{center}
    \includegraphics[width=0.95\textwidth]{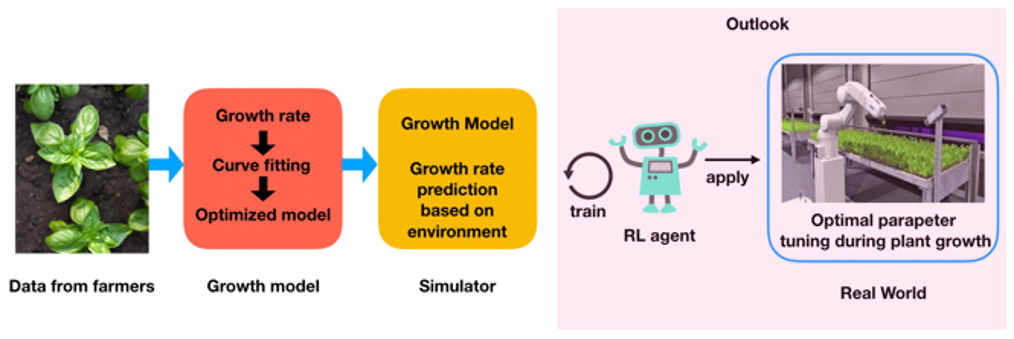}
  \end{center}
  \caption{Overview of our proposed plant growth modeling chain.}
  \label{fig:modelingChain}
\end{figure}
The corresponding modeling chain is illustrated in Fig. \ref{fig:modelingChain}. It comprises four main steps, (1) image processing, (2) fitting of growth curves, (3) learning the dependence of growth parameters on environment parameters, and (4) construction of the simulator. All steps are described in detail below. 

\subsection{Recording and processing of plant growth data} 

\paragraph{Dataset collection} 
The data used in this experiment has been collected over a period of 12 months of growing multiple batches of basil. The collection of data consists of two parts:
\begin{enumerate}
    \item Environmental data (using IoT data); 
    \item Top-down images of the crop (using cameras).
\end{enumerate}

In terms of (1) environmental data, the following variables were collected from the growth chamber every 5 minutes:
\begin{itemize}
    \item $\mathrm{CO}_2$
    \item Air temperature
    \item Relative air humidity
    \item Electrical Conductivity (EC)
    \item Oxidation Reduction Potential (ORP)
    \item Water pH
    \item Water temperature
\end{itemize}

To collect (2) top-down images of the crop, we used an 8-megapixel Raspberry Pi Camera Module 2. Crop images have been captured every 15 minutes. This low-cost camera module offers high-quality images and is suitable to deploy to both greenhouses and vertical farms to capture the growth of crops.

\paragraph{Processing} 
Upon collecting the dataset, we post-processed the images to quantify crop growth. Quantifying crop growth is done by estimating the size of the crop. This estimation is an object recognition problem consisting of two parts:
\begin{itemize}
    \item Crop localization; and
    \item Crop classification.
\end{itemize}
 
To be able to localize the crops, a potential approach is to remove the background using simple  thresholding. However, as the environment does not remain identical (e.g., crop grows, dimming of lights), simple thresholding yields unsatisfactory results when compared to statistical models.
Ideally, one needs to be able to accurately localize and classify the crop regardless of size and or unfavorable illumination conditions.

In contrast to the status quo, our approach uses deep convolutional neural networks (CNN) applied to a less researched area – plant growth in controlled environment agriculture. The reasoning behind our choice is that CNNs are considered state-of-the-art in image recognition \cite{gomez_towards_2016} thanks to their high learning capacity and robustness to the typical challenges of object recognition. To segment and classify tens of thousands of top-down images of basil, we fine-tuned a CNN using transfer learning\cite{aneja_transfer_2019}. The result is an accurate method to quantify the crop's size.

\subsection{Growth curves} 
To quantify the plant growth over time, we use the green canopy cover (CC) 
as a measure of plant size. Its development can be modeled with high average precision by the logistic growth curve \cite{vanuytrecht_aquacrop_2014}: 

\begin{equation}
CC(t)=\frac{CC_{\mathrm{max}}}{1+e^{-k(t-t_0)}}
\label{eq:growthCurve}
\end{equation} 
where $CC_{\mathrm{max}}$ is the maximum plant area, 
$t_0$ the midpoint, with $CC(t_0)=CC_{\mathrm{max}}/2$, and $k$ the plant growth rate. 

The data obtained from the image processing, i.e., the percentage area covered by plants, was simplified by averaging daily measurements. The reason for averaging daily measurements is to smooth out daily growth fluctuations due to light presence or absence. Individual batches were then fitted using \verb+SciPy+'s \verb|curve_fit| function \cite{2020SciPy-NMeth}. Lower and upper bounds were set to a fixed value of 0 and 1, respectively. The values 
$k$ and $t_0$ were obtained from fitting the model. The fitting was done continuously for all batches with a growth period of $D$ days and sub-batches starting from day 5 ($t=5$). 
 
In addition, the growth rates computed for all sub-batches can be used to iteratively estimate plant sizes at a future time step. This was only done for $t+1$ 
because the environmental parameters need to be adjusted at $t+1$. However, this adjustment can be extended further into the future. It is important to note that the estimates at this time are based solely on past plant sizes and do not take into account environmental factors. 

\begin{figure}
  \begin{center}
    \includegraphics[width=0.65\textwidth]{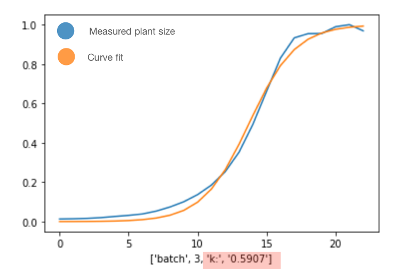}
  \end{center}
  \caption{Fitting of the growth curve example, showing the measured plant area per day (blue) and the fitted curve (orange).}
  \label{fig:growthCurve}
\end{figure}

\subsection{Dependence of growth parameter on environment parameters} 
\label{sec:environmentmodel}

To extend the simple growth model to a function reflecting growth dependencies on environment parameters, a model was fitted using the growth rates $k$ as target and environment parameters as input variables. Since the environment is the driving factor for plant growth, it is hypothesized that these variables can be used to predict $k$ or the change in 
$k$ for the next time step ($t+1$). 

\begin{figure}
  \begin{center}
    \includegraphics[width=0.45\textwidth]{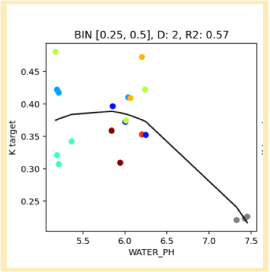}
  \end{center}
  \caption{Dependency of growth parameter k on water pH. Different colors correspond to different plant batches.}
  \label{fig:envDependence}
\end{figure}

For interpretability purposes, multiple polynomial regression models were fitted to reflect dependencies of growth rates (k) on individual environment parameters. To find optimal polynomial degrees a range of values [1, 2, 3, 4] has been tested. To improve the models relative to the model trained over the entire dataset, we binned the data into different growth stages. That is, we grouped the data into 4 bins of 25\% of total area covered by plants. 
This suggests that optimal environmental factors change during the growth process. 
Figure \ref{fig:envDependence} depicts such a dependency of $k$ on the pH level of water in the second quarter (25\% to 50\% of the final plant size). 
Fastest growth is obtained for pH values between 6.0 and 6.5, while smaller and especially higher values tend to slow down the growth process on average.

\subsection{Model training} 
The training set for the simulator-model consists of Basil batches. 
Given our initial dataset (30 batches * ~30 days), classic machine-learning methods were applied first. A complete list of tested models can be found in Table \ref{tab:modelperformance} below. The main input features consist of all environment variables from the current day ($d_{t}$), and the target variable, either  
$k$ or $\Delta k$, for the next day ($d_{t+1}$). Other feature combinations, such as $k * d_{t}$ and moving averages from two to six days, were included in the test iterations. 

To account for different growth stages, e.g., early growth or saturation, two different binning methods of the training data were tested in addition to training with the entire data set:  
\begin{enumerate}
    \item 4 sequential bins containing 25\% of the data, as described in Section \ref{sec:environmentmodel}.  
    \item 15 bins containing 40\% of the data with overlapping windows.  
\end{enumerate}

These bins were restricted with a lower bound of 0.05 and an upper bound of 1. This leads to smaller bins at the beginning and the end of the growth process, loosely reflecting the S-shape of the growth curve.

The dataset was then split into 80\% training data and 20\% test data. The validation was done using $R^2$ and MSE\cite{wiki_Coefficient_of_determination}. The best model for the Basil data regarding the MSE was achieved using a linear regression model with overlapping bins for predicting $k$ directly. The best model fit in terms of $R^2$ was achieved by a linear regression with sequential bins. The model for the simulator was chosen in terms of the smallest MSE.

\begin{figure}
  \begin{center}
    \includegraphics[width=0.65\textwidth]{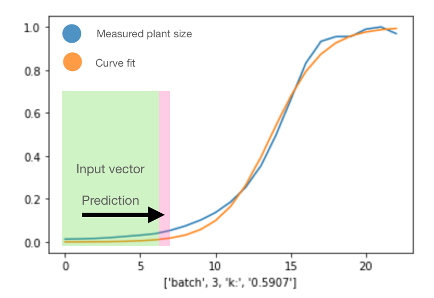}
  \end{center}
  \caption{Workings of the prediction model. Past observations of environment parameter settings and plant growth are used as input to predict next day's growth.}
  \label{fig:predictionModel}
\end{figure}

\begin{table}
  \caption{Performance comparison for different growth prediction model variants}
  \label{tab:modelperformance}
  \centering
      \begin{tabular}{llll}
    \toprule
    Model & Binning & $R^2$ & MSE \\
    \midrule
    Huber Regressor & overlapping &  0.704392 &   0.008783 \\
    Huber Regressor &  sequential &  0.758954 &   0.007161 \\
    Huber Regressor &  none &  0.503647 & 0.005309 \\
    Linear Regression &  overlapping &  0.988588 & \textbf{0.000354} \\
    Linear Regression &  sequential  &  \textbf{0.989152} & 0.001132 \\
    Linear Regression & none &  0.988169 & 0.000487 \\
    Polynomial Regressor & overlapping &  0.733265 & 0.009973 \\
    Polynomial Regressor & sequential &  0.792810 & 0.006454 \\
    Polynomial Regressor & none &  0.629016 & 0.014098 \\
    K-Nearest Neighbors &  overlapping &  0.848751 & 0.004200 \\
    K-Nearest Neighbors & sequential &  0.795600 & 0.005350 \\
    K-Nearest Neighbors & none &  0.911697 & 0.003541 \\
    Support Vector Machine &  overlapping &  0.681001 & 0.006655 \\
    Support Vector Machine & sequential &  0.638616 & 0.006822 \\
    Support Vector Machine &  none &  0.863949 & 0.003802 \\
    \bottomrule
    \end{tabular}
\end{table}

\subsection{Plant growth simulator}

The trained models for the various bins were implemented and tested in a simulation framework. Each growth batch was initialized with $k=0.2$ and $t_0=20$. The duration of the simulation was set to 35 days. At each step during the simulation, random environmental variables were drawn from a normal distribution with the mean and variance of each variable in the training set. 

During the simulation, the model of the bin whose center is closest to the current plant size is selected for the prediction of $k_{t+1}$. The obtained prediction for 
$k_{t+1}$ was then used to calculate 
$\Delta k$ by subtracting the current value, i.e. the last prediction of $k$, from this prediction: 

\begin{equation}
    \Delta \hat{k} =  \hat{k}_{t+1} - k_{t}   
\end{equation}

Further, $\Delta k$ was normalized to accommodate for potential changes in the environment variables during the simulation. Since sudden changes in the environment may lead to unrealistically large predictions of $\Delta k$, the following normalization has been used: 

\begin{equation}
    \Delta \hat{k}_{norm} = \Delta \hat{k}  \frac{\Delta k_{max}}{k_{max}}
\end{equation}

\begin{figure}
  \begin{center}
    \includegraphics[width=0.65\textwidth]{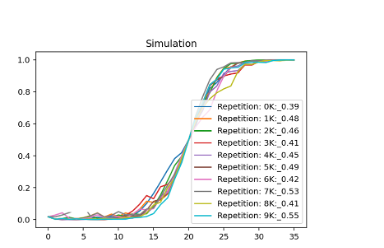}
  \end{center}
  \caption{Example plant growth simulations for different environmental conditions.}
  \label{fig:growthSimulations}
\end{figure}

After adding $\Delta \hat{k}_{norm}$ to $k_{t}$ the next day plant size can be calculated by inserting the value into the growth curve (Eq. \ref{eq:growthCurve}). Figure \ref{fig:growthSimulations} depicts 10 repetitions and the final growth rate $k$ of each simulated batch. 

To guide the system towards a desired target growth rate $k_{\mathrm{target}}$, we define the reward function as: 
\begin{equation}
   R = -100 \times (k - k_{\mathrm{target}})^2
\end{equation}
In our initial tries, we use the maximal value of $k$ observed in the data as the target value $k_{\mathrm{target}} = k_{max}$.

\begin{figure}
  \begin{center}
    \includegraphics[width=0.55\textwidth]{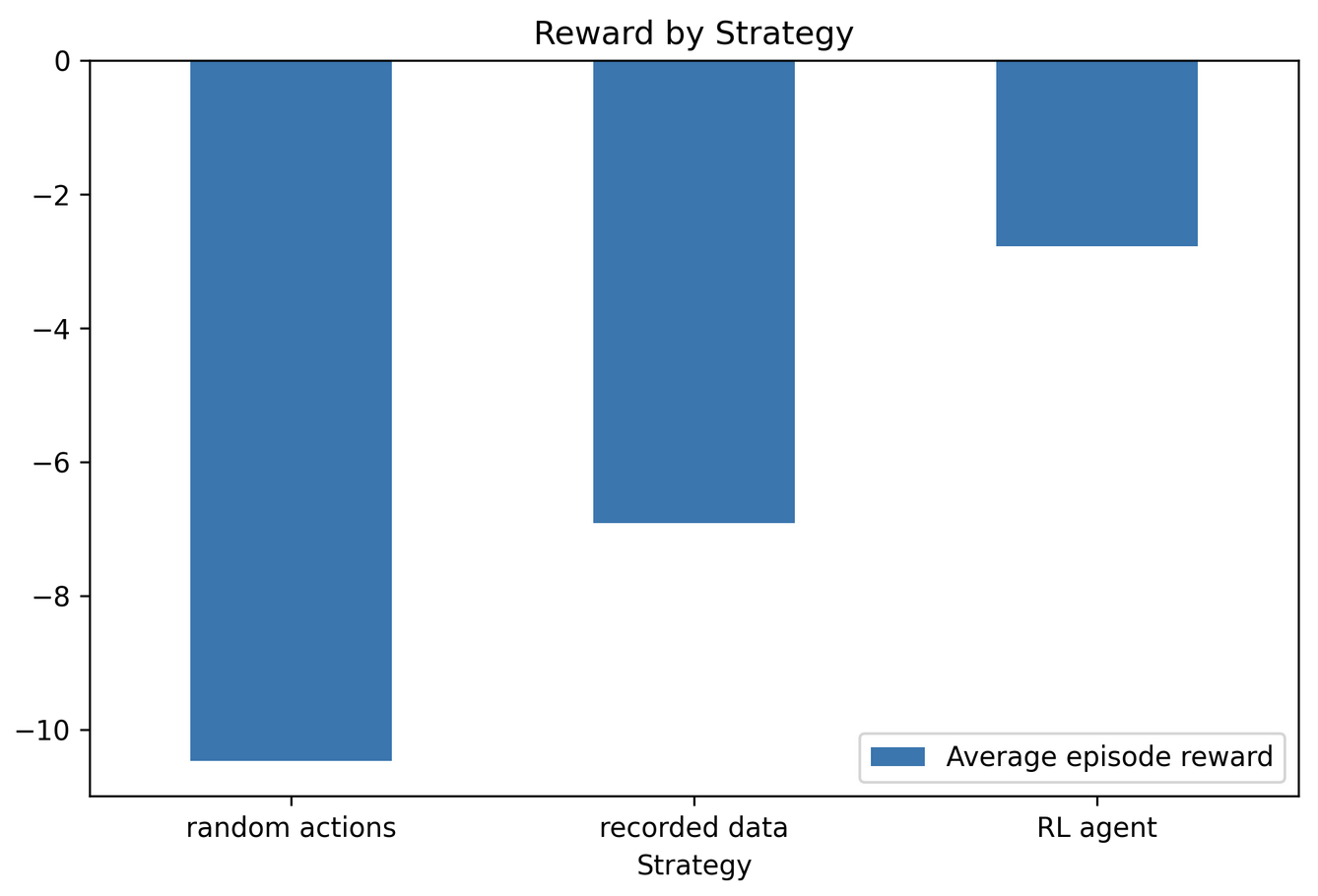}
  \end{center}
  \caption{Comparison of rewards achieved by different strategies (higher is better). The given values are the mean episode rewards achieved in a test environment over 1000 episodes, where one episode covers a full plant growth cycle from 3\% to 95\% of maximal plant area.}
  \label{fig:performance}
\end{figure}

We compared the interaction with the simulation environment of three different strategies and observed the rewards they achieve:
\begin{itemize}
    \item Random adjustments of plant environment parameters within the range of values observed in the data. 
    \item Applying the same adjustments that are present in the data. 
    \item Applying an AI-agent trained with PPO \cite{schulman_proximal_2017}.
\end{itemize}
The results are shown in Fig. \ref{fig:performance}. While the actions performed in the recorded data are certainly favorable to achieve faster plant growth we see a clear potential for AI agents trained in this way to improve outcomes in indoor farming environments.  

\section{Related Work} 
Plant growth models aim to describe the interaction between the growth of plants and their environment \cite{kang_plant_nodate}.
Several types of plant growth models exist, with varying degrees of complexity, centered on different key physiological processes and investigated at different temporal and spatial scales \cite{fourcaud_plant_2007}. 
An alternative is to learn the input-output relationships directly from data with the help of Machine Learning (ML), which has successfully been applied to plant growth prediction in the past. Conventional ML algorithms as well as deep learning were investigated, with LSTM \cite{hochreiter1997long} models giving the best results \cite{alhnaity_using_2020}. 
However, these models only learn to predict the regular temporal pattern of plant growth,
without providing insights into the effect of environmental parameters\cite{wang_predicting_2022}.
The usefulness of reinforcement learning to steer plant behavior has been recognized before and used in the construction of several game-like simulation environments \cite{hitti_growspace_2021,overweg_cropgym_2021}.
Recently, RL solutions for sustainable agriculture have been proposed \cite{binas_reinforcement_nodate},
and the first smart agriculture IoT systems based on reinforcement learning have been constructed in China \cite{Bu2019ASA,zhou_intelligent_2020}.
However, to date, no ML model exists which is able to actively control the complete growth process in an indoor farming setup, while taking into account growth speed, plant health and energy efficiency.

\section{Conclusions}
We developed a simulator for the automatic adjustment of environmental variables in indoor farming. We first fitted a growth curve model and subsequently trained a machine learning model to predict growth rates based on environmental variables. The resulting model could successfully be used in a simulator which enables the training of a reinforcement learning agent. Initial tests show potential for such agents to automate the control of plant environment parameters and optimize plant growth.
Next, we plan to incorporate plant health objectives as well as energy efficiency objectives in the reward function and test our AI agent in a real-world setting. 

\section*{Acknowledgements} 
We thank Innosuisse for the support within 59624.1 INNO-ICT.



\bibliographystyle{plain}    
\bibliography{library} 

\end{document}